\documentclass[showpacs,preprintnumbers,amsmath,amssymb,axodraw]{revtex4}

\usepackage{amsmath}
\usepackage{graphicx}
\usepackage{epsf}
\usepackage{psfrag}
\usepackage{epsfig}
\usepackage{graphics}
\usepackage[dvips]{color}

\begin{document}

\newcommand{\be}{\begin{equation}}
\newcommand{\ee}{\end{equation}}
\newcommand{\bq}{\begin{eqnarray}}
\newcommand{\eq}{\end{eqnarray}}

\title{Aspects of quantum corrections in a Lorentz-violating extension of the abelian Higgs model}

\date{\today}

\author{L. C. T. Brito$^{(a)}$} \email[]{lcbrito@dex.ufla.br}
\author{H. G. Fargnoli$^{(a)}$} \email[]{helvecio@dex.ufla.br}
\author{A. P. Ba\^eta Scarpelli$^{(b)}$} \email[]{scarpelli.apbs@dpf.gov.br}

\affiliation{(a) Universidade Federal de Lavras - Departamento de Ci\^encias Exatas \\
Caixa Postal 3037, 37.200-000, Lavras, Minas Gerais, Brazil}

\affiliation{(b)Setor T\'ecnico-Cient\'{\i}fico - Departamento de Pol\'{\i}cia Federal \\
Rua Hugo D'Antola, 95 - Lapa - S\~ao Paulo}

\begin{abstract}

\noindent
We investigate new aspects related to the abelian gauge-Higgs model with the addition of the Carroll-Field-Jackiw term. We focus
on one-loop quantum corrections to the photon and Higgs sectors due to spontaneous breaking of
gauge symmetry and show that new finite and definite Lorentz-breaking terms are induced. Specifically in the gauge sector, a CPT-even aether term is induced. Besides, aspects of the one-loop renormalization of the background vector dependent
terms are discussed.
\end{abstract}


\maketitle

\section{Introduction}

Lorentz and CPT symmetries play a fundamental role in Quantum Field theories and are observed with high precision
in all experimental tests (see \cite{data-exp} for a collection of experimental results).
However, the interest in Lorentz and CPT-violating models  has increased since a
Chern-Simons-like term in four dimensions was first considered \cite{jackiw}.
The Standard Model Extension (SME) \cite{kostelecky1}-\cite{coleman2}, which includes this Carroll-Field-Jackiw (CFJ) term, provides a description of
Lorentz and CPT violation, controlled by a set of coefficients whose small magnitudes are to be constrained by experiments. Many aspects
of the SME have been analyzed since then \cite{CS1}-\cite{andrade}.

Concerning the gauge sector, the SME encompasses two terms: the CFJ term, which has a Chern-Simons-like form and is CPT-odd, and a CPT-even one,
which is controlled by a constant fourth-rank background tensor with the same symmetries of the Riemann tensor. In the context of
the SME, there have been great interest  in the radiative induction of the CFJ term \cite{CS1}-\cite{CS15}. On the other
hand, the radiative generation of a particular form of the CPT-even term \cite{carroll}
has been studied in effective models which include Lorentz violating nonminimal couplings \cite{mgomes1}, \cite{mgomes2}, \cite{scarp-NM1}, \cite{scarp-NM2}.
One-loop corrections to the photon sector from a Lorentz-violating fermionic part have been performed in \cite{splitting}, in which it was obtained a non-zero amplitude for the vacuum triple photon splitting.

Spontaneous gauge symmetry breaking (SSB) has also been studied in Lorentz-breaking models. In \cite{scarp-belich}, the CFJ term was added to the traditional
abelian gauge-Higgs model. Some interesting physical properties at tree level have been analyzed as, for example, the peculiarities in the way the
degrees of freedom are distributed amongst the physical modes of the theory after spontaneous gauge symmetry breaking takes place. The same procedure has been
adopted in the non-abelian case \cite{nonab}. SSB in a Lorentz-violating theory
was also analyzed in a model which includes CPT-even terms both in gauge and Higgs sectors \cite{Altschul}.
In this paper we investigate aspects of quantum corrections to the abelian Lorentz and CPT-violating gauge-Higgs model of \cite{scarp-belich}.
The one-loop renormalization of the terms which depend on the Lorentz-violating parameter is carried out. Besides, we show that as a consequence of
SSB new Lorentz-violating terms, which are finite, are radiatively induced both in gauge and Higgs sectors. These calculations, carried out after 
SSB takes place, involve massive photons, which have been recently studied in \cite{mass-photons} for Lorentz-violating models. 

The paper is organized as follows: in section II we present the model and its Feynman rules, section III is dedicated to the one-loop
renormalization of the background vector dependent terms, we calculate the finite radiatively induced terms in section IV and, finally,
the concluding comments are presented in section V.

\section{The model}

Let us first review the abelian gauge-Higgs model, defined by the classical
lagrangian density
\begin{equation}
\mathcal{L}_{GH}=-\frac{1}{4}F_{\mu\nu}F^{\mu\nu}+D_{\mu}\phi\left(D^{\mu}\phi\right)^{*}+\mu^{2}\phi\phi^{*}-\frac{\lambda}{4}\left(\phi\phi^{*}\right)^{2},
\label{eq:higgs_model}
\end{equation}
where $\lambda$ and $\mu^{2}$ are positive constants, the covariant
derivative is defined by $D_{\mu}\phi=\partial_{\mu}\phi+ieA_{\mu}\phi$
and $F_{\mu\nu}=\partial_{\mu}A_{\nu}-\partial_{\nu}A_{\mu}$ is the
electromagnetic field strength tensor. The lagrangian is invariant
under local $U(1)$ gauge transformations and the complex scalar field
$\phi$ develops a vacuum expectation value $\langle\phi\rangle_{0}=\frac{v}{\sqrt{2}}$
(with $v$ constant), since $U(1)$ symmetry is spontaneously broken.
The quantum field theory obtained from (\ref{eq:higgs_model}) is
renormalizable to all orders of perturbation theory \cite{renorm}.

Rewriting the lagrangian (\ref{eq:higgs_model}) in terms of real
scalar fields $\rho$ and $\varphi$, such that $\phi=2^{-\frac{1}{2}}\left(\rho+v+i\varphi\right)$,
yields
\begin{eqnarray}
\mathcal{L}_{GH}+\mathcal{L}_{GF}+\mathcal{L}_{ghost}&=&-\frac{1}{4}F_{\mu\nu}F^{\mu\nu}+\frac{m_{A}^{2}}{2}A^{\mu}A_{\mu}-\frac{1}{2\xi}\left(\partial_{\mu}A^{\mu}\right)^{2}\nonumber\\
&&+\partial_{\mu}\bar{c}\partial^{\mu}c-\xi m_{A}^{2}\bar{c}c-\frac{\xi}{v}m_{A}^{2}\bar{c}c\rho\nonumber\\
&&+\frac{1}{2}\left(\partial_{\mu}\rho\partial^{\mu}\rho+\partial_{\mu}\varphi\partial^{\mu}\varphi\right)-\frac{m_{\rho}^{2}}{2}\rho^{2}-\frac{m_{\varphi}^{2}}{2}\varphi^{2}\nonumber\\
&&-\frac{\lambda}{16}\left(\rho^{2}+\varphi^{2}\right)^{2}-\frac{\lambda}{4} v\rho\left(\rho^{2}+\varphi^{2}\right)\nonumber\\
&&+eA_{\mu}\left[\rho\partial^{\mu}\varphi-\varphi\partial^{\mu}\rho\right]+\frac{e^{2}}{2}A^{\mu}A_{\mu}\left[\rho^{2}+\varphi^{2}+2v\rho\right]\nonumber\\
&&-\frac{1}{2}\delta m^{2}\left(\rho^{2}+\varphi^{2}\right)-v\delta m^{2}\rho,\label{eq:higgs_model-1}
\end{eqnarray}
where $\delta m^{2}=-\mu^{2}+\lambda v^{2}$, $m_{\rho}^{2}=\frac{\lambda v^{2}}{2}$,
$m_{\varphi}^{2}=\xi m_{A}^{2}$ and $m_{A}^{2}=v^{2}e^{2}$.  In  (\ref{eq:higgs_model-1}) we have added to $\mathcal{L}_{GH}$ a gauge fixing term
$\mathcal{L}_{GF}=-\left(2\xi\right)^{-1}\left(\partial_{\mu}A^{\mu}-\xi v e\varphi\right)^{2}$ and the ghost contributions $\mathcal{L}_{ghost}$, which
couple the Higgs boson $\rho$ to the Fadeev-Popov ghost field
$c$.

Since $\langle\phi\rangle_{0}\neq0$, the vacuum expectation value
$\langle\rho\rangle_{0}$ of the field $\rho$ should vanish at the
classical level, that is, $\delta m^{2}=0$. This gives the $\rho$
field a mass $m_{\rho}$, whereas the $\varphi^{2}$ term has
a gauge dependent mass $m_{\varphi}$. It is well known that in the
quantum theory the field $\varphi$ is the Goldstone boson. Actually,
we can fix $\delta m^{2}$ as a counterterm to each order of perturbation
theory using the normalization condition
\begin{equation}
\langle\rho\rangle_{0}=0 \label{eq:normcond1}
\end{equation}
at some renormalization scale.

Now we turn to the model that we will consider in the following:
\begin{equation}
\mathcal{L}=\mathcal{L}_{GH}+\mathcal{L}_{GF}+\mathcal{L}_{ghost}+\frac{1}{2}\left(k_{AF}\right)^{\mu}\epsilon_{\mu\nu\rho\sigma}A^{\nu}F^{\rho\sigma}.\label{eq:CFJterm}
\end{equation}
The last term in the above lagrangian is the Lorentz-violating CPT-odd
Carroll-Field-Jackiw (CFJ) term, which will be treated here as a perturbation
and has  the   Feynman rule

\begin{center}
\raisebox{-0.3cm}{
\begin{minipage}{5 cm}
\includegraphics[height=0.40cm]{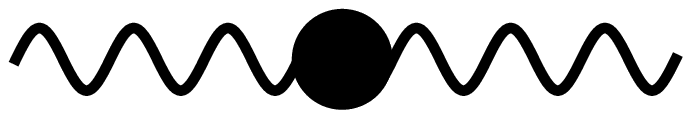}\qquad
\end{minipage}}
\begin{minipage}{4cm}\begin{equation}\label{CFJvertex} \ 2\epsilon_{\mu\nu\rho\sigma}\left(k_{AF}\right)^{\mu}p^{\rho}.\hfill \end{equation}\end{minipage}
\end{center}

The other relevant vertices defined from (\ref{eq:higgs_model-1}) are shown in the appendix. The propagators of the fields are given by:

\noindent $\bullet$ Photon propagator:
\begin{center}
\raisebox{-0.3cm}{\includegraphics[height=0.35cm]{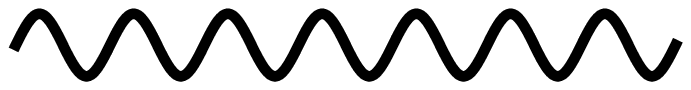}}\qquad
\begin{minipage}{7cm}\[
D^{\mu\nu}(p)=-\frac{i}{p^{2}-m_{A}^{2}}\left[\eta^{\mu\nu}+\left(\xi-1\right)\frac{1}{\left(p^{2}-\xi m_{A}^{2}\right)}p^{\mu}p^{\nu}\right];
\]
\end{minipage}
\end{center}

\noindent $\bullet$ Higgs propagator:
\begin{center}
\raisebox{-0.1cm}{\includegraphics[height=0.05cm]{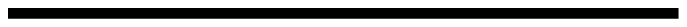}}\qquad
\begin{minipage}{3cm}
\[
\Delta_{\rho}(p)=\frac{i}{p^{2}-m_{\rho}^{2}};
\]
\end{minipage}
\end{center}

\noindent $\bullet$ Goldstone propagator:
\begin{center}
\raisebox{-0.1cm}{\includegraphics[height=0.05cm]{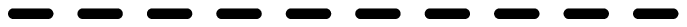}}\qquad
\begin{minipage}{3cm}
\[
\Delta_{\varphi}(p)=\frac{i}{p^{2}-m_{\varphi}^{2}};
\]
\end{minipage}
\end{center}

\noindent $\bullet$ Ghost propagator:
\begin{center}
\raisebox{-0.1cm}{\includegraphics[height=0.2cm]{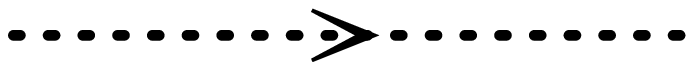}}\qquad
\begin{minipage}{3cm}
\[
\Delta_{c}(p)=\frac{i}{p^{2}-\xi m_{A}^{2}}.
\]
\end{minipage}
\end{center}

We are going to show explicitly in the next section that the one-loop divergences that arise from the CFJ perturbation are renormalized by the counterterm  $\delta m^2$.

\section{One-loop renormalization}\label{secaoii}

We will restrict our attention throughout this paper to one loop order in perturbation theory.
As we said before, the renormalization of the model (\ref{eq:CFJterm}) in the limit $\left(k_{AF}\right)^{\mu}\rightarrow0$
is well known. Hence in this section we will concentrate on quantum effects which arise from the CFJ term.
In other words, we will concentrate in diagrams with insertions of the ($AA$) vertex (\ref{CFJvertex}).

In addition to the $AA$ vertex, the lagrangian (\ref{eq:CFJterm})
has other four super-renormalizable vertices: $AA\rho$, $\rho^{3}$, $\bar{c}c\rho$ and $\rho\varphi^{2}$. Thus the superficial degree of divergence $d(G)$ of a diagram $G$ reads
\begin{equation}
d(G)=4-n-\left(V_{AA}+V_{AA\rho}+V_{\rho^{3}}+V_{\bar{c}c\rho}+V_{\rho\varphi^{2}}\right),\label{eq:div_superf}
\end{equation}
where $n$ is the number of external lines of $G$ and, in parentheses,
we have the sum of the number of super-renormalizable
vertices. From (\ref{eq:div_superf}), we can see that diagrams that
have $n=4$ and at least one super-renomalizable vertex are finite;
the same is true with $n=3$ and at least two super-renomalizable vertices.

\begin{figure}[ht]
\begin{center}
\includegraphics [height=3 cm]{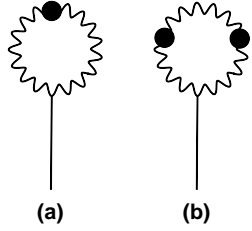}
\end{center}
\caption{Potentially divergent diagrams of the Higgs field one-point function. (a) One $AA$ vertex; (b) two $AA$ vertices.}
\label{umponto}
\end{figure}

\begin{figure}[ht]
\begin{center}
\includegraphics [height=2.1 cm]{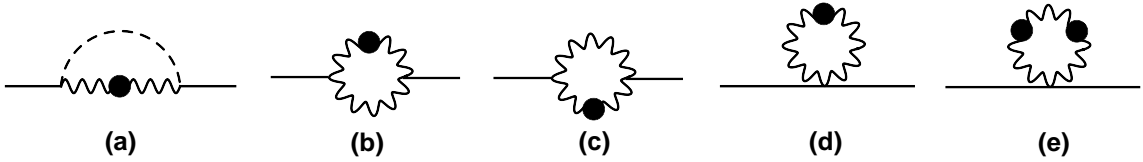}
\end{center}
\caption{Higgs two-point function divergent diagrams. (a), (b), (c) and (d) Diagrams with one $AA$ vertex; (e) diagram with two $AA$ vertices.}
\label{proparho}
\end{figure}

\begin{figure}[ht]
\begin{center}
\includegraphics [height=1.9 cm]{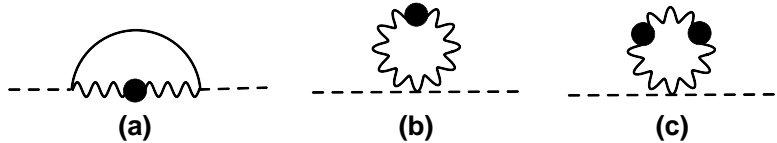}
\end{center}
\caption{Goldstone two-point function divergent diagrams. (a) and (b) Diagrams with one $AA$ vertex; (c) two $AA$ vertices.}
\label{propaphi}
\end{figure}

\begin{figure}[ht]
\begin{center}
\includegraphics[height=2.8 cm]{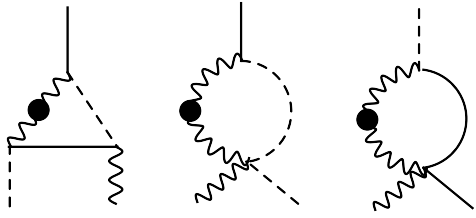}
\end{center}
\caption{Divergent vertex diagrams proportional to $\left(k_{AF}\right)_{\mu}$.}
\label{trespontos}
\end{figure}

\begin{figure}[ht]
\begin{center}
\includegraphics [height=1.2 cm]{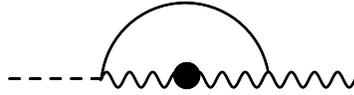}
\end{center}
\caption{Mixing propagator divergent diagram.}
\label{misto}
\end{figure}

Actually, the one loop diagrams in figures \ref{umponto}-\ref{misto} have $AA$ vertices and
$d(G)\geq0$. We will show in this section that the normalization condition (\ref{eq:normcond1}) - which determines the counterterm $\delta m^{2}$ -
is sufficient to renormalize the model at one loop.

We start our analysis with the $\rho$ field one-point function. The divergent diagrams (proportional to powers of
$\left(k_{AF}\right)_{\mu}$) are depicted in figure \ref{umponto} (in the following figures, external lines are amputated).

The expression of the diagram proportional to $\left(k_{AF}\right)_{\mu}$ is null. After using Feynman rules in an arbitrary gauge
and Dimensional Reduction \cite{dimred} (only the integrals are calculated in $d=4-\varepsilon$ dimensions),
the second diagram in figure \ref{umponto} gives the divergent term
\begin{equation}
\Gamma^{(div)}_{\rho} = -\frac{3i}{4\pi^2}ve^2\left(k_{AF}\right)^2\frac{1}{\varepsilon}.
\end{equation}

Thus, to obey the renormalization condition (\ref{eq:normcond1}) the divergent piece of $\delta m^2$ must be chosen as
\begin{equation}\label{escolha}
\delta m^2 = -\frac{3}{4\pi^2}e^2\left(k_{AF}\right)^2\frac{1}{\varepsilon}.
\end{equation}

We proceed next to the two-point functions. The gauge field propagator receives no divergent contribution (at one-loop) from the $AA$ vertex, as
we discuss in the next section. Figures \ref{proparho} and \ref{propaphi} show the superficially
divergent $\left(k_{AF}\right)_{\mu}$ contributions to the Higgs and Goldstone propagators, respectively.

Terms with $p_{\alpha}p_{\beta}\epsilon^{\mu\nu\alpha\beta}$ appear in the expressions of diagrams 2(a)-(d), 3(a) and 3(b) and then they are equal to
zero. The divergent pieces of diagrams 2(e) and 3(c) are
\be
\Gamma^{(div)}_{\rho\rho}=-\frac{3i}{4\pi^2}e^2\left(k_{AF}\right)^2\frac{1}{\varepsilon}\label{divrho}
\ee
and
\be
\Gamma^{(div)}_{\varphi\varphi}=-\frac{3i}{4\pi^2}e^2\left(k_{AF}\right)^2\frac{1}{\varepsilon}.\label{divphi}
\ee

Besides the one and two-point functions considered above, we must also evaluate the vertex function shown in figure \ref{trespontos},
because it has $n=3$ and equation (\ref{eq:div_superf}) tells us that those diagrams are superficially logarithmically divergent.
However, one can verify that the divergent piece of each diagram in figure \ref{trespontos} is individually null.
Therefore, no vertex function has divergences proportional to powers of the parameter $\left(k_{AF}\right)_{\mu}$ at one-loop order.

Lastly, the mixing two-point function $<\varphi A_{\mu}>$ has also logarithmic divergences. The $\left(k_{AF}\right)_{\mu}$ divergent contribution to it is given
by the diagram of figure \ref{misto}. Once more, it can be shown that the divergent piece of this diagram is null.

Thus, as we can see in  the lagrangian density (\ref{eq:higgs_model-1}), the choice (\ref{escolha}) is sufficient to remove the
divergences (\ref{divrho}) and (\ref{divphi}) rendering a one-loop finite theory.

\section{Finite quantum corrections to gauge and Higgs sectors}

\indent
We perform first the calculations of the one-loop corrections to the gauge self-energy. In addition to the usual gauge-Higgs diagram in figure 6, which arise from the spontaneous gauge symmetry breaking, we have  new diagrams with insertions of $AA$ vertex in the internal gauge
propagator. So we can obtain a perturbative series on $\left(k_{AF}\right)_\mu$, from which we will consider up to the second order in this parameter - figure 7.

\begin{figure}[ht]
\begin{center}
\includegraphics [height=1.5 cm]{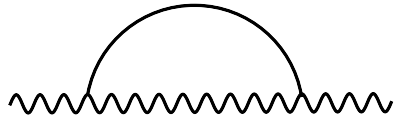}
\end{center}
\caption{Contribution due to the SSB.}
\label{propaphoton}
\end{figure}

\begin{figure}[ht]
\begin{center}
\includegraphics [height=1.5 cm]{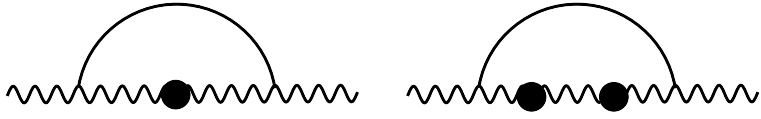}
\end{center}
\caption{Background vector dependent contributions to the photon propagator.}
\label{photonbola}
\end{figure}

The amplitude corresponding to the diagrams of figures \ref{propaphoton} and \ref{photonbola} is given by
\be
\Pi_{\mu \nu}^{\rho AA}=\Pi_{\mu \nu}^{(0)}+ \Pi_{\mu \nu}^{(1)}+ \Pi_{\mu \nu}^{(2)}+{\cal O}(\left(k_{AF}\right)^3),
\ee
where
\bq
&&\Pi_{\mu \nu}^{(0)}=-4v^2e^4 g_{\mu \nu}\int \frac{d^dk}{(2\pi)^d} \frac{1}{\left(k^2-m^2_A\right)\left[(p-k)^2-m^2_{\rho}\right]},
\eq
\bq
&&\Pi_{\mu \nu}^{(1)}=8iv^2e^4\left(k_{AF}\right)^{\lambda} \varepsilon_{\lambda\mu\rho\nu} \int \frac{d^4k}{(2\pi)^4}
\frac{k^\rho}{\left(k^2-m^2_A\right)^2\left[(p-k)^2-m^2_{\rho}\right]} \nonumber \\
&&= \frac{1}{2\pi^2}v^2e^4\left(k_{AF}\right)^{\lambda}p^{\rho}\varepsilon_{\lambda\mu\rho\nu}\int_{0}^{1}dx \frac{(1-x)x}{M^{2}_{A\rho}}
\eq
and
\bq
&&\Pi_{\mu \nu}^{(2)}=-16v^2e^4\int \frac{d^4k}{(2\pi)^4}
\frac{1}{\left(k^2-m^2_A\right)^3\left[(p-k)^2-m^2_{\rho}\right]} \times \nonumber\\
&& \left\{\left(k_{AF}\right)^2\left(k_{\mu}k_{\nu}-g_{\mu\nu}k^2\right)-\left(k_{AF}\right)\cdot k\left(k_{\mu}\left(k_{AF}\right)_{\nu}+k_{\nu}\left(k_{AF}\right)_{\mu}\right)+\left(k_{AF}\right)_{\mu}\left(k_{AF}\right)_{\nu}k^2
+(\left(k_{AF}\right)\cdot k)^2 g_{\mu\nu} \right\}\nonumber \\
&&= -\frac{iv^2e^4}{2\pi^2}\left\{\left[\left(k_{AF}\right)^2g_{\mu\nu}-\left(k_{AF}\right)_{\mu}\left(k_{AF}\right)_{\nu}\right]\int^{1}_{0}dx \frac{(1-x)^2}{M^2_{A\rho}} \right.\nonumber \\
&& \left.+ \left[\left(k_{AF}\right)^2\left(p_{\mu}p_{\nu}-g_{\mu\nu}p^2\right)+L_{\mu\nu}\right]\int_{0}^{1}dx \frac{(1-x)^2x^2}{M^4_{A\rho}}\right\}.\label{pidois}
\eq
We have defined above the expressions
\be
M^2_{12}\equiv m_1^2(1-x)+m_2^2x-p^2x(1-x)
\ee
and
\be
L_{\mu\nu} \equiv (\left(k_{AF}\right)\cdot p)^2g_{\mu\nu}-(\left(k_{AF}\right)\cdot p)\left(p_{\mu}\left(k_{AF}\right)_{\nu}+p_{\nu}\left(k_{AF}\right)b_{\mu}\right)+p^2\left(k_{AF}\right)_{\mu}\left(k_{AF}\right)_{\nu}.
\ee

Let us then analyze these three terms. The first one is the Proca-type correction which occur even in the usual gauge-Higgs model.
The second one, $\Pi_{\mu \nu}^{(1)}$, will give a finite correction to the CFJ term. The most interesting term is $\Pi_{\mu \nu}^{(2)}$.

Analyzing the structures present on (\ref{pidois}), we see that the first term on brackets violates gauge and Lorentz symmetries and
the piece $\left(k_{AF}\right)^2\left(p_{\mu}p_{\nu}-g_{\mu\nu}p^2\right)$ is a correction to the Maxwell term with dependence on the
Lorentz-violating parameter $\left(k_{AF}\right)_\mu$. The part on $L_{\mu \nu}$ is a
CPT-even Lorentz breaking contribution. The induced term in the Lagrange density has the following aether-like form \cite{carroll}:
\be
{\cal L}_{aether}= v^2e^4B(p,m_{\rho},m_A)\left( k_{AF}^\mu F_{\mu \nu} \right)^2
\ee
with $B(p,m_{\rho},m_A) \equiv \frac{i}{4\pi^2}\int_{0}^{1}dx \frac{(1-x)^2x^2}{M^4_{A\rho}}$. This term can be mapped in the CPT-even term proposed in \cite{kostelecky2},
\be
{\cal L}_{gauge}^{even}=-\frac 14 \left(k_{F}\right)_{\mu \nu \alpha \beta} F^{\mu \nu} F^{\alpha \beta},
\ee
as long as we establish the relation
\bq
&&\left(k_{F}\right)_{\mu \nu \alpha \beta}= -v^2e^4B(p,m_{\rho},m_{A}) \left\{
g_{\mu \alpha} \left(k_{AF}\right)_\nu \left(k_{AF}\right)_\beta - g_{\nu \alpha} \left(k_{AF}\right)_\mu \left(k_{AF}\right)_\beta \right. \nonumber \\
&&\left. + g_{\nu \beta} \left(k_{AF}\right)_\mu \left(k_{AF}\right)_\alpha - g_{\mu \beta} \left(k_{AF}\right)_\nu \left(k_{AF}\right)_\alpha
 \right\}.
\eq

We now consider the background vector dependent quantum corrections to the Higgs field propagator. Figures 2 and 8 give contributions up to second order in $\left(k_{AF}\right)_\mu$.

\begin{figure}[ht]
\begin{center}
\includegraphics [height=1.3 cm]{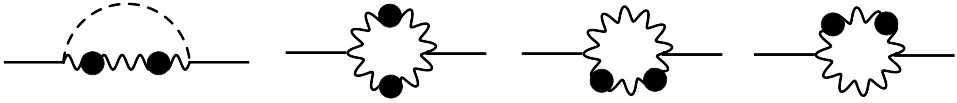}
\end{center}
\caption{Remaining contributions to $\rho$ propagator due to the Lorentz violating term of order $\mathcal{O}(\left(k_{AF}\right)^2)$.}
\label{duasinsercoes}
\end{figure}

As already mentioned in section \ref{secaoii}, diagrams 2(a)-(d) vanish. Summing up the diagrams depicted in figures 2(e) and \ref{duasinsercoes}, we have
\begin{eqnarray}
&&\Gamma_{\rho\rho} = \frac{i}{4\pi^2}e^2\left\{-\frac{3}{2}\left(k_{AF}\right)^2\left[\frac{2}{\varepsilon}
-\gamma+\ln(4\pi)-\frac{5}{3}-\ln\left(\frac{m^2_{A}}{\Lambda^2}\right)\right]
+\frac{3}{2}v^2e^2\left(k_{AF}\right)^2\left[2\int_{0}^{1}dx\frac{(1-x)x}{M^{2}_{AA}}\right.\right.\nonumber\\
&&\left.\left.+\int_{0}^{1}dx\frac{(1-x)^2}{M^2_{AA}}\right]+\left[(p\cdot \left(k_{AF}\right))^2-\left(k_{AF}\right)^2p^2\right]
\left[2\int_{0}^{1}dx \frac{(1-x)^2}{M^2_{AA}}-v^2e^2\int_{0}^{1}dx \frac{(1-x)^2x^2}{M^4_{AA}}\right]\right\},
\end{eqnarray}
where $\gamma$ is the Euler-Mascheroni constant and $\Lambda$ is an energy scale.

The first term in brackets is the divergent piece appearing in section \ref{secaoii} plus some constants,
the second is a mass correction and the last one can be mapped in
\begin{equation}
\mathcal{L}_{Higgs}^{even}=\frac{1}{2}\kappa^{\mu\nu}\partial_{\mu}\rho\partial_{\nu}\rho,
\end{equation}
introduced in \cite{kostelecky2}, provided that
\begin{equation}
\kappa^{\mu\nu}=2e^2\left(\left(k_{AF}\right)^2g^{\mu\nu}-\left(k_{AF}\right)^{\mu}\left(k_{AF}\right)^{\nu}\right)A(p,m_A),
\end{equation}
with
\be
A(p,m_A)=\frac{i}{4\pi^2}\left[2\int_{0}^{1}dx \frac{(1-x)^2}{M^2_{AA}}-v^2e^2\int_{0}^{1}dx \frac{(1-x)^2x^2}{M^4_{AA}}\right].
\ee

A comment is in order. We have shown that, at second order in the Lorentz-violating parameter $\left(k_{AF}\right)_\mu$, finite and well defined CPT-even terms are induced both in the photon and the Higgs sectors. These inductions are consequences of the spontaneous breaking of local $U(1)$ symmetry, which produces  the $\rho AA$ vertex. This is an interesting theoretical result, although the magnitude of the background vector $\left(k_{AF}\right)$ has been constrained to the level of $10^{-42}$, as can be found in \cite{data-exp}. Concerning this small magnitude of the Lorentz-breaking parameter, it is worth to mention that, although $|\left(k_{AF}\right) e^4|> |\left(k_{AF}\right)^2 e^2|$, which would imply that two-loop contributions should be considered, these $e^4$ linear terms in the background vector would not contribute to the CPT-even part.

Last but not least, particularly intriguing is the induction of a gauge-breaking term in the Lorentz-violating sector of the model. A question which arises for a further investigation is whether this term
(and the other radiatively induced terms) would be canceled out
if the finite part of the one-loop shift in the Higgs field is carried out.

\section{Concluding  comments}

\indent
We have studied some aspects of quantum corrections of the four-dimensional abelian Lorentz-violating CFJ-Higgs model.
The new divergences due the presence of the CFJ term were shown to be worked out by the renormalization condition which requires
the vacuum expectation  value of the Higgs field remains null. We have also analyzed what kind of effects are
induced at the quantum level by spontaneous gauge symmetry breaking due the presence of the Chern-Simons-like term. It was shown that,
at second order in the background vector, it is induced an aether-like CPT-even Lorentz breaking term in the pure gauge sector. This
term is finite and free from ambiguities. A finite and determined CPT-even term is also induced in the pure Higgs sector.

It is worth to comment on the induction of a finite gauge and Lorentz breaking term in the photon sector. It is possible that, after considering
the finite part of the one-loop shift in the Higgs field, this term would be canceled out.
A further investigation would be interesting in order to check this  and  also whether the other radiatively induced terms could be canceled
out in this process.

\begin{center}\textbf{Acknowledgements}\end{center}

H. G. Fargnoli and A. P. Ba\^eta Scarpelli thank CNPq for the financial support. A. P. Ba\^eta Scarpelli thanks J. A. Helayel-Neto for relevant discussions. H.G. Fargnoli thanks S.M. de Souza for his encouragement.

\section*{Appendix}

We present in this appendix the relevant vertex Feynman rules derived from lagrangian (\ref{eq:higgs_model-1}).
\vspace{0.3cm}

\noindent $\bullet$ $AA\rho$ vertex:
\begin{center}
\raisebox{-1.2cm}{\includegraphics[height=2cm]{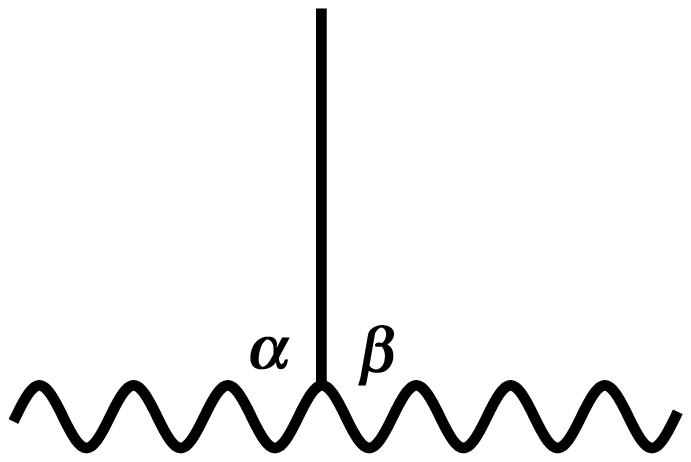}}
\begin{minipage}{5cm}\[ive^2 g_{\alpha\beta}\]
\end{minipage}
\end{center}

\noindent $\bullet$ $A\rho\phi$ vertex:
\begin{center}
\raisebox{-1.2cm}{\includegraphics[height=2cm]{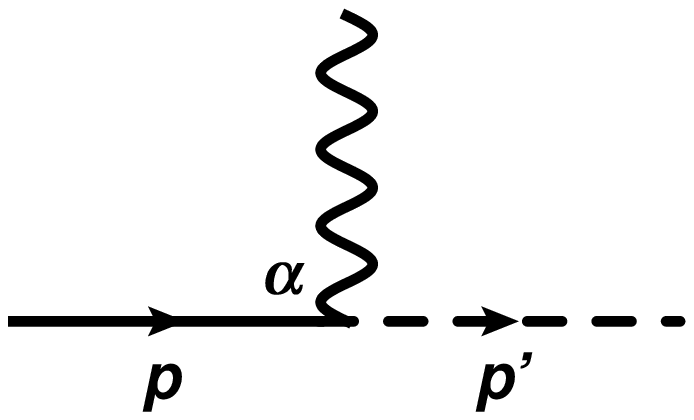}}
\begin{minipage}{5cm}\[e\left(p + p'\right)_{\alpha}\]
\end{minipage}
\end{center}

\noindent $\bullet$ $AA\rho\rho$ vertex:
\begin{center}
\raisebox{-1.3cm}{\includegraphics[height=2.2cm]{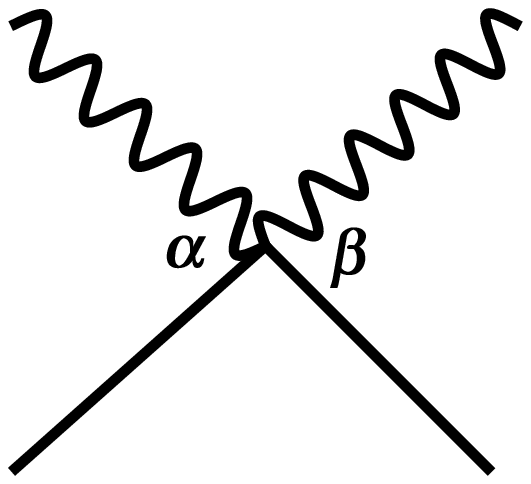}}
\begin{minipage}{5cm}\[ie^2g_{\alpha\beta}\]
\end{minipage}
\end{center}

\noindent $\bullet$ $AA\phi\phi$ vertex:
\begin{center}
\raisebox{-1.3cm}{\includegraphics[height=2.2cm]{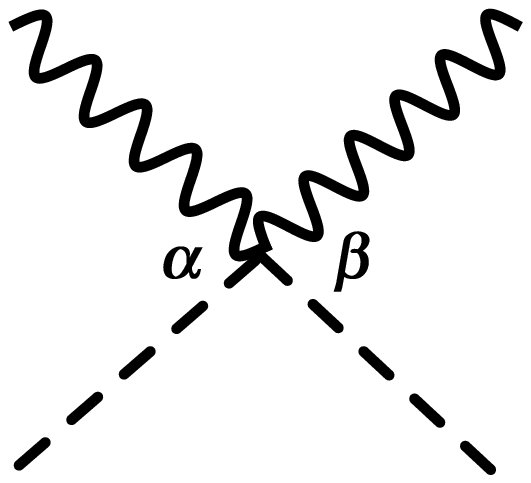}}
\begin{minipage}{5cm}\[ie^2g_{\alpha\beta}\]
\end{minipage}
\end{center}

\end{document}